\documentclass[prl,twocolumn,showpacs,preprintnumbers,floats,floatfix,superscriptaddress,amsmath,amssymb]{revtex4}

\usepackage{graphicx}
\usepackage{dcolumn}
\usepackage{bm}
\def\bra#1{\mathinner{\langle{#1}|}}
\def\ket#1{\mathinner{|{#1}\rangle}}

\begin{document}

\title{Witness for initial system-environment correlations in open system dynamics}

\author{Elsi-Mari Laine}\email{emelai@utu.fi}
\affiliation{Turku Centre for Quantum Physics, Department of Physics and Astronomy, University of
Turku, FI-20014 Turun yliopisto, Finland}

\author{Jyrki Piilo}\email{jyrki.piilo@utu.fi}
\affiliation{Turku Centre for Quantum Physics, Department of
             Physics and Astronomy, University of Turku, FI-20014 Turun yliopisto, Finland}

\author{Heinz-Peter Breuer}\email{breuer@physik.uni-freiburg.de}
\affiliation{Physikalisches Institut, Universit\"at Freiburg, Hermann-Herder-Strasse 3,
             D-79104 Freiburg, Germany}

\date{\today}

\begin{abstract}
We study the evolution of a general open quantum system when the
system and its environment are initially correlated. We show that
the trace distance between two states of the open system can
increase above its initial value, and derive tight upper bounds
for the growth of the distinguishability of open system states.
This represents a generalization of the contraction property of
quantum dynamical maps. The obtained inequalities can be
interpreted in terms of the exchange of information between the
system and the environment, and lead to a witness for
system-environment correlations which can be determined through
measurements on the open system alone.
\end{abstract}


\maketitle

When the initial state of an open quantum system is statistically
independent from its environment, the evolution of the reduced
system can be described by a family of completely positive
dynamical maps between the reduced system states. The most simple
evolution for initially uncorrelated states is described by a
Markov process for which this family of maps forms a dynamical
semigroup \cite{Gorini,Lindblad}. However, in many systems the
Markov description gives an overly simplified picture of the
dynamics and a more rigorous treatment is needed \cite{Breuer}.
Many methods for treating non-Markovian dynamics have been
developed in recent years \cite{nm
jumps,breuer2,breuer3,daffer,Kossakowski}, but the effect of
initial correlations is often ignored. The assumption of initially
uncorrelated states is well justified whenever the system and the
environment are weakly interacting, but it has been argued
\cite{ic1} that the assumption of initially uncorrelated states
generally is too restrictive. Therefore, the influence of initial
correlations on the open system dynamics has been recently under
an intensive study \cite{ic1,ic2,ic3,ic4,ic5,ic6,ic7,lidar}.

If one considers initial states of the total system with different
system-environment correlations, the open system dynamics can in
general no longer be described by a dynamical map acting on the
reduced state space \cite{ic1,ic2,ic4}. This is a common physical
situation which occurs, for example, when the initial correlations
are created by an earlier interaction between the system and its
environment. The question thus arises, whether in this situation
one can still find general quantitative features that characterize
the reduced system dynamics. The answer to this question is, in
fact, yes: We demonstrate below that the distinguishability
between any two states of the open system can increase above its
initial value. This increase has a tight upper bound which can be
interpreted as the relative information on the initial states of
the total system which is inaccessible for the open system, i.~e.,
which cannot be obtained through measurements on the open system
at the initial time. The existence of this upper bound can be seen
as a generalization of the contraction property of dynamical maps
and is shown to lead to a measurable witness for correlations in
the initial system-environment states.

In the following we shall use the trace distance as a distance
measure for quantum states. The trace distance of two quantum
states $\rho^1$ and $\rho^2$ is defined as
$D(\rho^1,\rho^2)=\frac{1}{2}\textrm{Tr}|\rho^1-\rho^2|$. It is a
metric on the space of physical states, satisfying $0\leq D\leq
1$, and represents the achievable upper bound for the
distinguishability between the probability distributions arising
from measurements performed on the quantum states \cite{nielsen}.
Thus, the trace distance can be given an interpretation as the
distinguishability between two quantum states.

Consider now a bipartite quantum system consisting of a system $S$
coupled to an environment $E$, such that together they form an
isolated system described by the initial state $\rho_{SE}$. The
state of the system $S$ at time $t$ can then be written as
\begin{equation}
 \rho_S(t) = \textrm{Tr}_E\big[U_t\rho_{SE}U^\dagger_t\big],
 \label{QDP}
\end{equation}
where $U_t=\exp[-iHt/\hbar]$ represents the unitary time evolution
operator of the composite system with total Hamiltonian $H$, and
${\textrm{Tr}}_E$ denotes the partial trace over the environment.
Given two initial states $\rho_{SE}^1$ and $\rho_{SE}^2$ of the
composite system the rate of change of the trace distance between
the corresponding reduced system states $\rho_S^1(t)$ and
$\rho_S^2(t)$ at time $t$ is given by
\begin{equation}
 \sigma(t) = \frac{d}{dt}D(\rho_S^1(t),\rho_S^2(t)).
 \label{rate}
\end{equation}
For $\sigma(t)<0$ the trace distance and the
distinguishability between the reduced states decrease. We
interpret this as a flow of information from the system to the
environment. Correspondingly, whenever we have $\sigma(t)>0$ the
distinguishability of the pair of reduced states increases, and we
interpret this increase of distinguishability as a reversed flow
of information from the environment to the system \cite{nmprl}.

If one assumes that the system and the environment are initially
uncorrelated with a fixed environmental state $\rho_E$, i.~e.,
$\rho_{SE}=\rho_S\otimes\rho_E$, one can describe the time
evolution of the reduced system given by Eq.~\eqref{QDP} through a
family of completely positive dynamical maps $\Phi_t$,
\[
 \rho_S \mapsto \rho_S(t) = \Phi_t\rho_S
 = \textrm{Tr}_E\big[U_t\rho_S\otimes\rho_EU^\dagger_t\big],
\]
which map the state space of the reduced system into itself. It is
a well known fact \cite{nielsen,RUSKAI} that such dynamical maps
are contractions for the trace distance, i.~e., for any initial
pair of states $\rho_S^{1,2}$ and for any time $t\geq 0$ we have
\begin{equation}
 D(\Phi_t\rho_S^1,\Phi_t\rho_S^2)\leq D(\rho_S^1,\rho_S^2).
\label{contraction}
\end{equation}
Hence, for initially uncorrelated total system states and
identical environment states, the trace distance between the
reduced system states at time $t$ can never be larger than the
trace distance between the initial states, as is illustrated by
the two lower curves in Fig.~1(a). Physically this means that the
total amount of the information flowing back from the environment
to the system is bounded from above by the total amount of the
information earlier flowed out from the system since the initial
time zero.

In the presence of initial correlations the dynamics of the
reduced system given by Eq.~\eqref{QDP} can, in general, not be
described by a map acting on the open system's state space,
because two different total system states with one and the same
reduced state may evolve in time into states with different
reduced system states \cite{ic1,ic2,ic4}. We can still give the
quantity given in Eq. \eqref{rate} an interpretation in terms of
the flow of information between the system and the environment.
However, if initial correlations are present
Eq.~\eqref{contraction} does not apply. This allows the situation
where the trace distance of the reduced system states grows to
values which are larger than the initial trace distance as is
illustrated by the upmost curve in Fig.~1(a).

Our aim is to construct upper bounds for the growth of the trace
distance in the presence of initial correlations. To this end, we
consider an arbitrary pair of initial states $\rho_{SE}^{1,2}$ of
the total system with corresponding reduced system states
$\rho_S^{1,2}=\textrm{Tr}_E\big[\rho_{SE}^{1,2}\big]$ and
environment states
$\rho^{1,2}_E=\textrm{Tr}_S\big[\rho_{SE}^{1,2}\big]$. One then
finds the inequality
\begin{eqnarray}  \label{inequality-1}
 &&D\left(\textrm{Tr}_E\big[U_t\rho_{SE}^1U_t^\dagger\big],
 \textrm{Tr}_E\big[U_t\rho_{SE}^2U_t^\dagger\big]\right)-D(\rho_{S}^1,\rho_{S}^2) \nonumber \\
 &&\leq D(\rho_{SE}^1,\rho_{SE}^2)-D(\rho_{S}^1,\rho_{S}^2)
 \equiv I(\rho_{SE}^1,\rho_{SE}^2),
\end{eqnarray}
which states that the increase of the trace distance of the states
$\rho_S^1$ and $\rho_S^2$ during the time evolution given by
Eq.~\eqref{QDP} is bounded from above by the quantity
$I(\rho_{SE}^1,\rho_{SE}^2)$. The inequality can easily be derived
by employing the invariance of the trace distance under unitary
transformations and by using the fact that the trace distance is
non-increasing under the partial trace operation. The upper bound
$I(\rho_{SE}^1,\rho_{SE}^2)$ represents the distinguishability of
the initial states $\rho_{SE}^{1,2}$ of the total system minus the
distinguishability of the corresponding reduced system states
$\rho_S^{1,2}$. Hence, this quantity represents the loss of
distinguishability of the initial total states which results when
measurements on the reduced system only can be performed. One can
thus interpret $I(\rho_{SE}^1,\rho_{SE}^2)$ as the relative
information on the initial states which lies outside the open
system and is inaccessible for it. The inequality
\eqref{inequality-1} therefore leads to the following physical
interpretation: The maximal amount of information the open system
can gain from the environment is the amount of information flowed
out earlier from the system since the initial time, plus the
information which is initially outside the open system.

\begin{figure}
\includegraphics[width=0.5\textwidth]{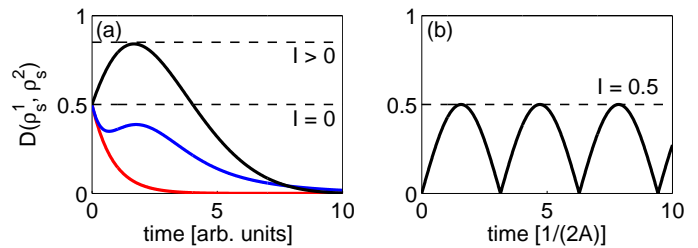}
\caption{(color online) (a) Schematic picture of the dynamics of
the trace distance. For $I(\rho_{SE}^1,\rho_{SE}^2)=0$ the trace
distance decreases monotonically according to a Markovian dynamics
[light gray (red) line], or shows a non-monotonic behavior in the
non-Markovian case [dark gray (blue) line], but can never exceed
the initial value marked by the lower dashed line. The black curve
illustrates the dynamics in a case with
$I(\rho_{SE}^1,\rho_{SE}^2)>0$ for which initially inaccessible
information allows the increase of the trace distance over the
initial value. The bound of Eq.~\eqref{inequality-1} is indicated
by the upper dashed line. (b) Exact trace distance dynamics with
initial correlations for the spin-bath model discussed in the
text.} \label{fig1}
\end{figure}

As we shall demonstrate in several examples below, the inequality
\eqref{inequality-1} can in fact become an equality, showing that
the upper bound of this inequality is tight. When the bound of
\eqref{inequality-1} is actually reached at a certain time $t$,
the distinguishability of the reduced system states at time $t$ is
equal to the distinguishability of the total system states at time
zero. This means that the relative information on the total system
states at time zero has been transferred completely to the reduced
system at time $t$.

Obviously, the contraction property \eqref{contraction} for
completely positive maps is a special case of the inequality
\eqref{inequality-1} which occurs if both total initial states are
taken to be uncorrelated with the same environmental state, i.~e.
$\rho_{SE}^{1,2}=\rho_S^{1,2}\otimes\rho_E$. This follows by using
the invariance of the trace distance under the tensor product
which yields $I(\rho_S^1\otimes\rho_E,\rho_S^2\otimes\rho_E)=0$,
implying that initially there is no information outside the
reduced system.

A further important special case of the inequality
\eqref{inequality-1}, which reveals most clearly the role of
initial correlations, is obtained if we choose $\rho_{SE}^2$ to be
the fully uncorrelated state constructed from the marginals of
$\rho_{SE}^1$, i.~e., $\rho_{SE}^2=\rho_S^1\otimes\rho^1_E$. In
this case inequality \eqref{inequality-1} simplifies to
\begin{eqnarray}
 && D\left(\textrm{Tr}_E\big[U_t\rho_{SE}^1U_t^\dagger\big],
 \textrm{Tr}_E\big[U_t\rho_S^1\otimes\rho^1_EU_t^\dagger\big]\right)\nonumber\\
 && \leq  D(\rho_{SE}^1,\rho_{S}^1\otimes\rho^1_{E}).
 \label{simple inequality}
\end{eqnarray}
This inequality shows how far from each other two initially
indistinguishable reduced states can evolve when only one of the
two total initial states is correlated. Here, the relative
information $I(\rho_{SE}^1,\rho_{SE}^2)$ on the total initial
states which is inaccessible for the open system is equal to
$D(\rho^1_{SE},\rho_S^1\otimes\rho^1_E)$. This quantity describes
how well the state $\rho_{SE}^1$ can be distinguished from the
corresponding fully uncorrelated state $\rho_S^1\otimes\rho^1_E$
and, therefore, provides a measure for the amount of correlations
in the state $\rho_{SE}^1$. Thus, the increase of the trace
distance is bounded from above by the correlations in the initial
state.

Returning to the general case, we use the subadditivity of the
trace distance with respect to tensor products,
\[
 D(\rho_S^1\otimes\rho_E^1,\rho_S^2\otimes\rho_E^2)
 \leq D(\rho_S^1,\rho_S^2) + D(\rho_E^1,\rho_E^2),
\]
to conclude from inequality \eqref{inequality-1}
\begin{eqnarray} \label{inequality-2}
 && D\left(\textrm{Tr}_E\big[U_t\rho_{SE}^1U_t^\dagger\big],
 \textrm{Tr}_E\big[U_t\rho_{SE}^2U_t^\dagger\big]\right)-D(\rho_{S}^1,\rho_{S}^2) \\
 && \leq D(\rho_{SE}^1,\rho_{SE}^2) - D(\rho_S^1\otimes\rho_E^1,\rho_S^2\otimes\rho_E^2)
 + D(\rho_E^1,\rho_E^2) \nonumber \\
 && \leq D(\rho_{SE}^1,\rho_S^1\otimes\rho_E^1) + D(\rho_{SE}^2,\rho_S^2\otimes\rho_E^2)
 + D(\rho_E^1,\rho_E^2), \nonumber
\end{eqnarray}
where the second inequality follows by using twice the triangle
inequality for the trace distance. This inequality clearly shows
that in the most general case an increase of the trace distance of
the reduced states implies that there are initial correlations in
$\rho_{SE}^1$ or $\rho_{SE}^2$, or that the initial environmental states
are different. In particular, an increase of the trace distance
can obviously occur when the environmental states are different,
even if the initial total states are uncorrelated. However, if the
environmental states are fixed then any increase of the trace
distance is a witness for the presence of initial correlations.

We discuss several examples to illustrate the inequalities derived
above. First, we consider two qubits under the controlled-NOT
gate, i.~e., under an interaction described by the unitary
operator $U_C:\ket{00}\to \ket{00}, \ket{01}\to
\ket{01},\ket{10}\to \ket{11},\ket{11}\to \ket{10}$. We regard the
first qubit as the system $S$ (control qubit), and the second
qubit as the environment (target qubit). Let us study the states
\cite{ic4}
\begin{eqnarray}
&&\rho_{SE}^1=(\alpha\ket{00}+\beta\ket{11})(\alpha^*\bra{00}+\beta^*\bra{11}),\nonumber\\
&&\rho_{SE}^2=|\alpha|^2\ket{00}\bra{00}+|\beta|^2\ket{11}\bra{11}
\label{states}
\end{eqnarray}
with $\alpha,\beta \neq 0$. The state $\rho_{SE}^1$ is a pure
entangled state, while $\rho_{SE}^2$ is a mixed state with only
classical correlations (it is separable and has zero quantum
discord \cite{Zurek, Vedral}). For these two total states the
system states and the environmental states coincide:
\[
 \rho_S^1=\rho_S^2=|\alpha|^2\ket{0}\bra{0}+|\beta|^2\ket{1}\bra{1}
 =\rho_E^1=\rho_E^2.
\]
Under the action of the controlled-NOT gate the increase of the
trace distance is found to be
\[
 D\left(\textrm{Tr}_E\big[U_{C}\rho_{SE}^1U_{C}^\dagger\big],
 \textrm{Tr}_E\big[U_{C}\rho_{SE}^2U_{C}^\dagger\big]\right)=|\alpha\beta|,
\]
witnessing that at least one of the initial states must have been
correlated. The right-hand side of the inequality
\eqref{inequality-1} is given by
$D(\rho_{SE}^1,\rho_{SE}^2)=|\alpha\beta|$ which shows that the
upper bound of this inequality is indeed reached here and that,
therefore, the information on the initial states of the total
system is transferred completely to the reduced system by the
controlled-NOT gate.

Let us also study a situation described by the inequality
\eqref{simple inequality} for a pair consisting of a correlated
and the corresponding uncorrelated state. For the state
$\rho_{SE}^1$ in Eq.~\eqref{states} we obtain
\[
 D\left(\textrm{Tr}_E\big[U_{C}\rho_{SE}^1U_{C}^\dagger\big],
 \textrm{Tr}_E\big[U_{C}\rho_{S}^1\otimes\rho_{E}^1U_{C}^\dagger\big]\right)=|\alpha\beta|,
\]
We also have
$D(\rho_{SE}^1,\rho_{S}^1\otimes\rho^1_{E})=|\alpha\beta|+|\alpha\beta|^2$
and, hence, the bound of \eqref{simple inequality} is not reached
for this case, but the increase of the trace distance still
witnesses the correlations in the initial state.

To give an example of a negative result for an initially
correlated state let us study the classically correlated state
$\rho_{SE}^2$ of Eq.~\eqref{states}. The trace distance between
the reduced states
$\textrm{Tr}_E\big[U_{C}\rho_{SE}^2U_{C}^\dagger\big]$ and
$\textrm{Tr}_E\big[U_{C}\rho_{S}^2\otimes\rho_{E}^2U_{C}^\dagger\big]$
is found to vanish. Hence, although $\rho_{SE}^2$ is correlated,
\begin{equation} \label{correlation}
 D(\rho_{SE}^2,\rho_{S}^2\otimes\rho_{E}^2)=2|\alpha\beta|^2,
\end{equation}
the trace distance between the system states does not increase.
However, if we first apply the controlled-NOT gate and then a swap
operation $U_{\textrm{swap}}$, we do obtain a growth of the trace
distance. The swap operation corresponds to taking the target
qubit as the system and the control qubit as the environment. One
finds
\begin{equation} \label{trace-distance}
 D\left(\textrm{Tr}_E\big[U\rho_{SE}^2U^\dagger\big],
 \textrm{Tr}_E\big[U\rho_{S}^2\otimes\rho_{E}^2U^\dagger\big]\right)=2|\alpha\beta|^2,
\end{equation}
where $U=U_{\textrm{swap}}U_C$. This example demonstrates that an
increase of the trace distance can occur even in the case in which
the initial states are purely classically correlated. Moreover, we
observe from Eqs.~\eqref{correlation} and \eqref{trace-distance}
that the equality sign in \eqref{simple inequality} holds,
demonstrating again the tightness of the bound provided by this
inequality.

As our final example we consider the model of a central spin with
Pauli operator $\bm{\sigma}$ which interacts with a bath of $N$
spins with Pauli operators $\bm{\sigma}^{(k)}$ through the
Hamiltonian
$H=A_0\sum_{k=1}^N{(\sigma_+\sigma_-^{(k)}+\sigma_-\sigma_+^{(k)})}$.
We investigate the initial states
\begin{eqnarray*}
 \rho_{SE}^1&=&\ket{\Psi}\bra{\Psi},\quad \ket{\Psi}=
 \alpha \ket{-}\otimes\ket{\chi_+}+\beta\ket{+}\otimes\ket{\chi_-}\\
 \rho_{SE}^2&=&|\alpha|^2\ket{-}\bra{-}\otimes\ket{\chi_+}\bra{\chi_+}
 +|\beta|^2\ket{+}\bra{+}\otimes\ket{\chi_-}\bra{\chi_-},\nonumber
\end{eqnarray*}
where $\ket{\pm}$ are central spin states, and
$\ket{\chi_+}=\ket{++\dots+}$ and
$\ket{\chi_-}=\frac{i}{\sqrt{N}}\sum_k\ket{k}$ are environment
states. Here, the state $\ket{k}$ is obtained from $\ket{\chi_+}$
by flipping the $k$th bath spin. The states $\rho_{SE}^{1,2}$ have
the same marginals and, thus, differ from one another only by the
correlations. We find that the increase of the trace distance is
given by
\begin{equation}
 D\left(\textrm{Tr}_E\big[U_t\rho_{SE}^1U_t^\dagger\big],
 \textrm{Tr}_E\big[U_t\rho_{SE}^2U_t^\dagger\big]\right)
 = | \Re(\alpha^*\beta) \sin(2At)|, \label{spinb}
\end{equation}
where $A=\sqrt{N}A_0$. The trace distance thus oscillates
periodically between the initial value zero and the maximal value
$|\Re(\alpha^*\beta)|$. We conclude that for almost all values of
the amplitudes $\alpha$ and $\beta$ there is an increase of the
trace distance witnessing the initial correlations. Moreover, we
have $D(\rho_{SE}^1,\rho_{SE}^2)=|\alpha\beta|$. Hence, if
$\alpha^*\beta$ is real the upper bound of inequality
\eqref{inequality-1} is reached periodically whenever
$|\sin(2At)|=1$, as is shown in Fig.~1(b) for the case
$\alpha=\beta=1/\sqrt{2}$.

In order to use the increase of the trace distance as a witness
for initial correlations in an experiment one has to make sure
that the environmental state is initially fixed. Provided that
this can be done, one has to measure whether the trace distance
increases over its initial value during the time evolution. If one
gets a positive result, i.e., if one finds that
$D\big(\textrm{Tr}_E\big[U_t\rho_{SE}^1U_t^\dagger\big],
\textrm{Tr}_E\big[U_t\rho_{SE}^2U_t^\dagger\big]\big)>D(\rho_{S}^1,\rho_{S}^2)$
for some instant of time, the inequality \eqref{inequality-2}
implies that at least one of the initial states was correlated.
From a negative result one cannot draw any conclusions about the
initial state; as we have seen there are situations in which there
are correlations in the initial state although the trace distance
does not increase over the initial value.

In summary, we have studied the dynamics of open systems with
correlations in the initial system-environment states. It has been
shown that the growth of the distinguishability of reduced system
states is bounded from above by the relative information on the
initial states lying outside the open system. The obtained
inequalities can be interpreted in terms of the exchange of
information between the system and its environment: If the trace
distance increases over its initial value, the information which
is initially inaccessible is transferred to the open system,
enlarging the distinguishability of its states. In our
presentation we do not assume the initial correlations to be
fixed, i.~e., we do not consider maps between the reduced states,
but rather transformations from the composite system states to the
reduced system states. The results obtained here also suggest ways
to experimentally observe the presence of initial correlations
which do not require any prior knowledge about the structure of
the environment and the system-environment interaction. Thus our
results do not only clarify the consequences of the presence of
initial correlations on the dynamics, but can be also useful in
designing experiments for the detection of initial correlations.

\acknowledgments We thank the Magnus Ehrnrooth Foundation, the
Academy of Finland (project 133682), and the Finnish National
Graduate School of Modern Optics and Photonics for financial
support.

\end{document}